\documentclass{natureprintstyle}

\usepackage{graphicx}
\usepackage{amsmath, amssymb}
\usepackage{multirow}
\usepackage{ifthen}
\usepackage{cite}

\newcommand{\capbf}[1]{\textbf{\textsf{#1}}}

\title{Shedding Light on Three-Body Recombination in an Ultracold Atomic Gas}

\author{A. H\"arter$^1$, A. Kr\"ukow$^1$, M. Dei\ss$^1$, B. Drews$^1$, E. Tiemann$^2$, \& J. Hecker Denschlag$^1$}

\begin{document}
\twocolumn[
\maketitle

\begin{affiliations}
  \item Institut f\"ur Quantenmaterie and Center for Integrated Quantum Science and Technology IQ$^\text
  {ST}$, Universit\"at Ulm, 89069 Ulm, Germany
  
  \item Institut f\"ur Quantenoptik, Leibniz Universit\"at Hannover, 30167 Hannover, Germany
\end{affiliations}

\begin{abstract}
Three-body recombination is a prime example of the fundamental
interaction between three particles. Due to the complexity of this
process it has resisted a comprehensive description. Experimental
investigations have mainly focussed on the observation of
corresponding loss rates without revealing information on the
reaction products. Here, we provide the first general experimental
study on the population distribution of molecular quantum states
after three-body recombination in a non-resonant regime. We have
developed a highly sensitive detection scheme which combines
photoionization of the molecules with subsequent ion trapping. By
analyzing the ionization spectrum, we identify the population of
energy levels with binding energies up to $h\times 750\:$GHz. We
find a broad population of electronic and nuclear spin states and
determine a range of populated vibrational and rotational states.
The method presented here can be expanded to provide a full survey
of the products of the recombination process. This may be pivotal
in developing an in-depth model that can qualitatively and
quantitatively predict the
reaction products of three-body recombination.\\

\end{abstract}
]

While cold collisions of two atoms are understood to an excellent
degree, the addition of a third collision partner drastically
complicates the interaction dynamics. In the context of
Bose-Einstein condensation in atomic gases, three-body
recombination plays a crucial
role\cite{Hess1983,Burt1997,Soding1999,Esry1999} and it
constitutes a current frontier of few-body
physics\cite{Suno2009,Wang2011,Guevara2012}. However, the
investigations focussed mainly on the atom loss rates established
by the recombination events. Discussions of the final states
populated in the recombination process were restricted to the
special case of large two-body scattering
lengths\cite{Fedichev1996, Bedaque2000} and culminated in the
prediction and observation of Efimov resonances\cite{Efimov1970,
Braaten2001, Kraemer2006}. In the limit of large scattering
lengths, recombination has been seen to predominantly yield
molecules in the most weakly bound states$\,$\cite{Weber2003,
Jochim2003a}. However, in the more general case
of a scattering length comparable to the van der Waals radius, the
recombination products might depend on details of the interaction
potential. In fact, ongoing theoretical studies using simplified
models indicate that recombination does not necessarily always
favor the most weakly bound state$\,$\cite{dIncao2013}(see also$\,$\cite{Simoni2006}).
In general, recombination processes are of fundamental interest in
various physical systems\cite{Bates1962, Hess1983, Flower2007}.
The control and tunability of ultracold atomic systems provide an
experimental testbed for a detailed understanding of the nature of
these processes.

\begin{figure}
\begin{center}
\vspace{6pt}
\includegraphics[width = 0.47\textwidth]{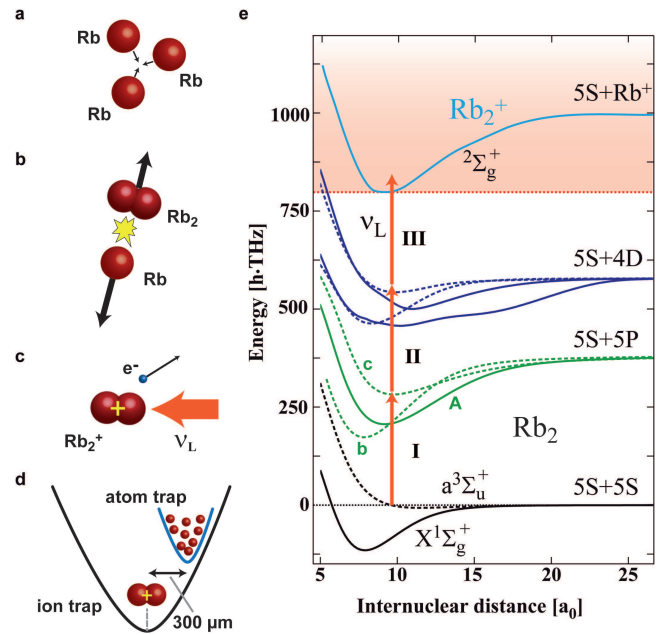}
\end{center}
\caption{\bf{Illustration of recombination and ionization in
the atom-ion trap. a}, A three-body collision in the ultracold gas
of $^{87}$Rb atoms leads to a recombination event in which,
\bf{b}, a Rb$_2$ molecule is formed with high kinetic energy.
 \bf{c}, While the atom is lost from the trap, the molecule can
 be photoionized in a REMPI process and trapped in the Paul trap. \bf{d},
 The relative positions of the atom and ion trap centers are shifted by about
$300\:\mu\textrm{m}$ to avoid atom-ion collisions. \bf{e},
Potential energy curves of the Rb$_2$ and Rb$_2^+$ molecule
adapted from refs \cite{Lozeille2006, Aymar2003}. The curves $A,
b, c,$  are $A^1\Sigma_u^+$, $b^3\Pi_u$, $c^3\Sigma_u^+$.
The internuclear distance is given in units of Bohr radii a$_0$. A REMPI
path with three photons is shown. It can create Rb$_2^+$ ions
in vibrational states up to $v\approx 17$.}
\label{fig:1}
\end{figure}

Here, we demonstrate the probing of molecules with binding
energies up to $h\times 750\:$GHz (where $h$ is Planck's constant)
generated via three-body
recombination of ultracold thermal $^{87}$Rb atoms. We produce
the atomic sample in an optical dipole trap located within a
linear Paul trap. The recombination and detection process is
illustrated in Fig.$\,$\ref{fig:1}a-d. Following a recombination
event, the created Rb$_2$ molecule can undergo resonance-enhanced
multi-photon ionization (REMPI) by absorbing photons from the
dipole trap laser at a wavelength around 1064.5$\:$nm. The ion is
then captured in the Paul trap and detected essentially background-free with very
high sensitivity on the single particle level. Fig.$\,$\ref{fig:1}e shows a simplified scheme
of the Rb$_2$ and Rb$_2^+$ potential energy curves. From
weakly-bound molecular states three photons suffice to reach the
molecular ionization threshold. An additional photon may
dissociate the molecular ion. By scanning the frequency of the
dipole trap laser by more than 60$\:$GHz we obtained a high
resolution spectrum featuring more than 100 resonance peaks. This
dense and complex spectrum contains information which vibrational,
rotational and hyperfine levels of the Rb$_2$ molecule are populated.
We present an analysis of these data and make a first assignment of the most
prominent resonances. This assignment indicates that in the recombination events
a broad range of levels is populated in terms of vibrational, rotational, electronic
and nuclear spin quantum numbers.

Our experimental scheme to detect cold molecules makes use of the
generally excellent detection efficiencies attainable for trapped
ions. It is related to proven techniques where
cold molecules in magneto-optical traps were photoionized from the
singlet and triplet ground states\cite{Fioretti1998, Gabbanini2000, Lozeille2006,
Huang2006, Salzmann2008, Sullivan2011} (see also
ref$\,$\cite{Mudrich2009}). Our method is novel
as it introduces the usage of a hybrid atom-ion trap which significantly
improves the detection sensitivity. We perform the following experimental
sequence. A thermal atomic sample typically containing
$N_\text{at}\approx 5 \times 10^5$ spin-polarized $^{87}$Rb atoms
in the $|F=1,m_\textrm{F}=-1\rangle$ hyperfine state is prepared
in a crossed optical dipole trap at a magnetic field of about
5$\:$G. The trap is positioned onto the nodal line of the
radiofrequency field of a linear Paul trap. Along the axis of the
Paul trap the centers of the atom and ion trap are separated by
about $300\:\mu\textrm{m}$ to avoid unwanted atom-ion collisions
(Fig.$\,$\ref{fig:1}d). At atomic temperatures of about
$700\:\textrm{nK}$ and peak densities $n_\text{0}\approx5\times
10^{13}\:\textrm{cm}^{-3}$ the total three-body recombination rate
in the gas is $\Gamma_\text{rec}=L_3 n_\text{0}^2
N_\text{at}/3^{5/2}\approx 10\:$kHz. Here, the three-body loss
rate coefficient $L_3$ was taken from ref$\,$\cite{Soding1999}. At
the rate $\Gamma_\text{rec}$, pairs of Rb$_2$ molecules and Rb
atoms are formed as final products of the reactions. Both atom and
molecule would generally be lost from the shallow neutral particle trap due to the
comparatively large kinetic energy they gain in the recombination
event (in our case typically on the order of a few K $\times
k_\text{B}$ where $k_\text{B}$ is the Boltzmann constant). The
molecule, however, can be state-selectively ionized in a REMPI
process driven by the dipole trap laser. All of these molecular
ions remain trapped in the deep Paul trap and are detected with
single particle sensitivity (see Methods). In each experimental
run, we hold the atomic sample for a time $\tau\approx10\:$s.
After this time we measure the number of produced ions in the trap
from which we derive (after averaging over tens of runs) the ion
production rate $\Gamma_\textrm{ion}$ normalized to a cloud atom
number of $10^6$ atoms.

As a consistency check of our assumption that Rb$_2$ molecules are ionized in the REMPI process, we verify the production of Rb$_2^+$
molecules. For this, we perform ion mass spectrometry in the Paul trap (see Methods). We detect primarily molecular Rb$_2^+$ ions,
a good fraction of atomic Rb$^+$ ions but no Rb$_3^+$ ions. Our experiments show that Rb$^+$ ions are produced in
light-assisted collisions of Rb$_2^+$ ions with Rb atoms on timescales below a few ms.
Details of this dissociation mechanism are currently under investigation and will be discussed elsewhere.

\begin{figure}
\begin{center}
\vspace{6pt}
\includegraphics[width = 0.4\textwidth]{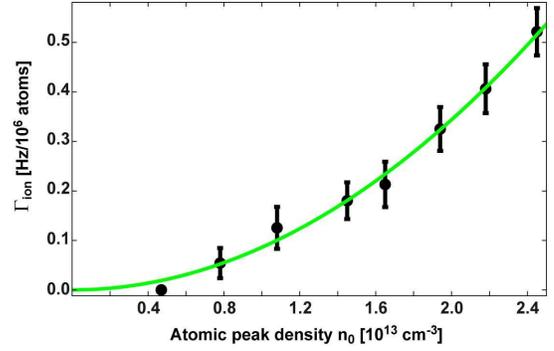}
\end{center}
\caption{\bf{Dependence of the ion production rate
$\Gamma_\textrm{ion}$ on atomic density.} $\Gamma_\textrm{ion}$
is normalized to a cloud atom number of $10^6$ atoms. The data are well
described by a quadratic fit (solid green line). They were taken
at a constant dipole trap laser intensity $I_\textrm{L}= 36
\:\textrm{kW}/\textrm{cm}^2$ and a laser frequency of
$\nu_\text{L}= 281630\:\textrm{GHz}$.
 }
\label{fig:2}
\end{figure}

Two pathways for the production of our neutral Rb$_2$ molecules
come immediately to mind. One pathway is far-off-resonant
photoassociation of two colliding Rb atoms (here with a detuning
of about 500$\,$GHz$\times h$). This pathway can be ruled out using several
arguments, the background of which will be discussed in more depth
later. For one, we observe molecules with a parity that is
incompatible with photoassociation of totally spin polarized
ensembles. Furthermore, we observe a
dependence of the ion production rate on light intensity that is
too weak to explain photoassociation.

The second pathway is three-body recombination of Rb atoms.
Indeed, by investigating the dependence of the ion production rate
$\Gamma_\textrm{ion}$ (which is normalized to a cloud atom number
of 10$^6$ atoms) on atomic density, we find the expected quadratic
dependence (see Fig.$\,$\ref{fig:2}).
For this measurement the density was adjusted by varying the cloud
atom number while keeping the light intensity of the dipole trap
constant.

\begin{figure*}
\begin{center}
\vspace{6pt}
\includegraphics[width = 0.73\textwidth]{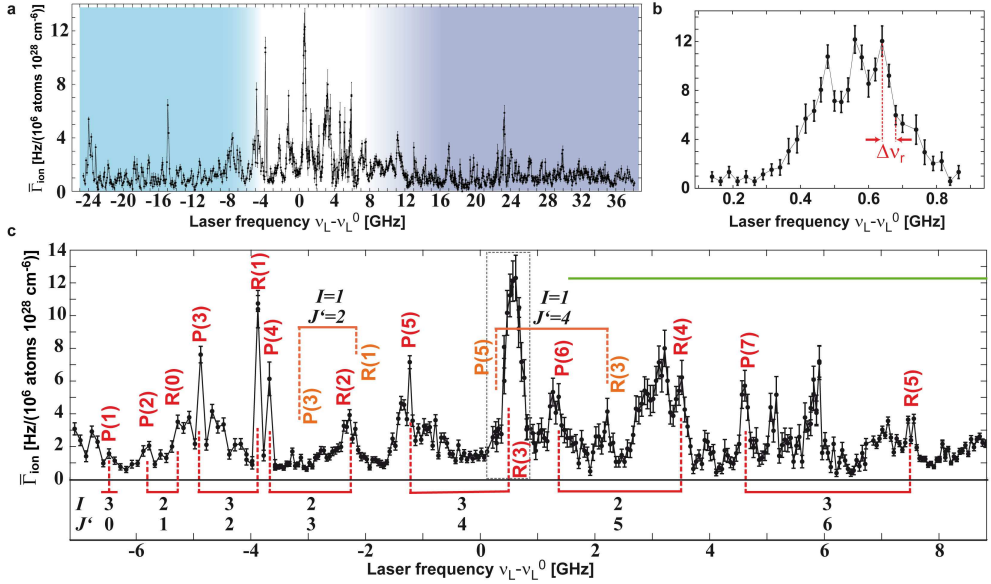}
\end{center}
\caption{\bf{REMPI spectrum. a}, A scan of the dipole trap laser frequency
$\nu_\text{L}$ over more than $60\:\textrm{GHz}$ around
an offset frequency $\nu_\text{L}^0=281.610\:\text{THz}$ shows
a multitude of resonance lines. Each data point is the result of
30 to 60 repetitions of the experiment with ion detection on the
single particle level. The total spectrum was obtained over a time
span of 2 months. Checks of the long-term consistency of resonance
positions and strengths were performed. Spectral regions dominated by
transitions to $c^3\Sigma_{g}^+$ are indicated by the shaded
areas in dark and light blue ($0_{g}^-$ and $1_{g}$ component, respectively).
\textbf{b}, High-resolution scan
of the strong resonance at $\nu_\text{L}-\nu_\text{L}^0\approx 0.5\:$GHz.
\textbf{c}, Central spectral region with assigned P/R branches
of the transition $X^1\Sigma_g^+(v=115) \rightarrow
A^1\Sigma_u^+(v'=68)$. The corresponding quantum numbers $I$ and $J'$ are given). P(J) marks
the transition $J\rightarrow J+1$, R(J) the transition
$J\rightarrow J-1$. These lines can be grouped into pairs sharing the same $J'$ of the excited state and $I$ quantum number. The region
where also transitions to $b^3\Pi_{u}$ appear is marked by a green horizontal bar.
}
\label{fig:3}
\end{figure*}

Next, we investigate the dependence of the ion production rate on the wavelength of the narrow-linewidth dipole trap laser (see Methods).
We scan the wavelength over a range of about 0.3$\:$nm around 1064.5$\:$nm, corresponding to a frequency range of about 60$\:$GHz. Typical frequency step sizes
are 50$\:$MHz or 100$\:$MHz. We obtain a rich spectrum of resonance lines which is shown in Fig.$\,$\ref{fig:3}a. The quantity $\bar{\Gamma}_\textrm{ion}$
denotes the ion production rate normalized to the atom number of the cloud and to the square of the atomic peak density.
We find strongly varying
resonance strengths and at first sight fairly irregular frequency spacings.
In the following we will argue that most resonance lines can be attributed to respective well-defined molecular levels
(resolving vibrational, rotational and often even hyperfine structure) that have been populated in the recombination process.
These levels are located in the triplet or singlet ground state,
$a^3\Sigma_u^+$ and $X^1\Sigma_g^+$, respectively.
The relatively dense distribution of these lines reflects that a fairly broad range of states is populated. A direct assignment
of the observed resonances is challenging,  as it hinges on the precise knowledge of the level structure of all the relevant ground and excited states.
In the following we will access and understand the data step by step.

One feature of the spectrum that catches the eye is the narrow
linewidth of many lines. For example, Fig.$\,$\ref{fig:3}b shows a
resonance of which the substructures have typical half-widths
$\Delta\nu_r\approx50\:\textrm{MHz}$. This allows us to roughly
estimate the maximal binding energy of the molecules involved.
Since the velocity of the colliding ultracold atoms is extremely
low, the kinetics of the recombination products is dominated by
the released molecular binding energy $E_\textrm{b}$. Due to
energy and momentum conservation the molecules will be expelled
from the reaction with a molecular velocity $v_\textrm{Rb2}=
\sqrt{2 E_\textrm{b}/(3 m_\textrm{Rb2})}$ where $m_\textrm{Rb2}$
is the molecular mass. The molecular resonance frequency $\nu_0$
will then be Doppler-broadened 
with a half-width $\Delta \nu_\textrm{D}=\sqrt{3}\nu_0
v_\textrm{Rb2}/2c$. Here, $c$ is the speed of light. By comparing
$\Delta \nu_\textrm{D}$ to the observed values of $\Delta\nu_r$ we
estimate a maximal binding energy on the order of
$E_\textrm{b,max}\approx h\times 2.5\:$THz. This simple analysis
overestimates the value $E_\textrm{b,max}$ since it neglects the
natural linewidth of the transition and possible saturation
broadening. Still, it already strongly constrains the possible
populated molecular levels that are observed in our experiment.

\begin{figure*}
\begin{center}
\vspace{6pt}
\includegraphics[width = 0.8\textwidth]{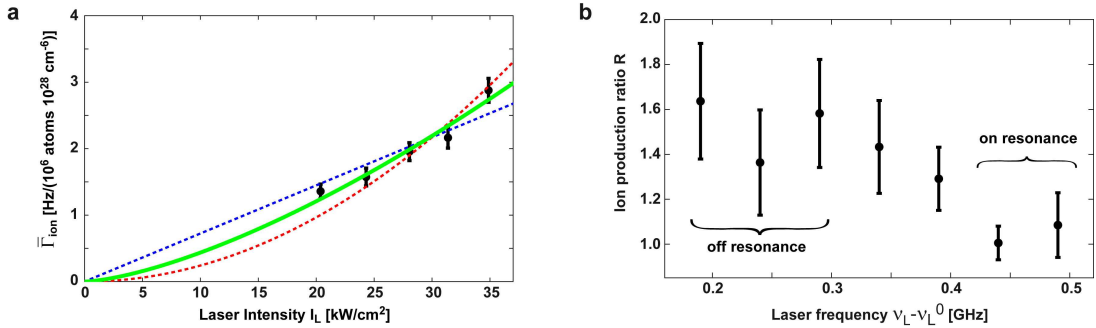}
\end{center}
\caption{\bf{Dependence of the ion production rate on the
intensity of the dipole trap laser. a}, Assuming a power-law
dependence $\bar{\Gamma}_\textrm{ion}\propto I_\textrm{L}^\alpha$,
the best fit to the data is achieved for $\alpha\approx 1.5$ (solid
green line). Linear and quadratic fits are also given (blue and
red dashed lines, respectively). \textbf{b}, Measurement of the
intensity dependence using a "chopped" dipole trap.
The ratio $R \approx 1$ on resonance indicates saturation of both
transitions I and II.
 }
\label{fig:4}
\end{figure*}

Next, we investigate the dependence of the ion production rate on laser intensity $I_\textrm{L}$.
In our experimental setup, this measurement is rather involved because the laser driving the REMPI process also confines the
atomic cloud. Thus, simply changing only the laser intensity would undesirably also change the density $n_\text{0}$ of the atoms. To prevent this
from happening we keep $n_\text{0}$ constant ($n_\text{0}\approx 5\times 10^{13}\:\textrm{cm}^{-3}$) by adjusting
the atom number and temperature appropriately. Due to these experimental complications we can only vary $I_\textrm{L}$ roughly by a factor of 2
 (Fig.$\,$\ref{fig:4}a).
We set the laser frequency to the value of $\nu_\text{L}=\nu_\text{L}^0 \equiv 281610\:\textrm{GHz}$, on the tail
of a large resonance (see Fig.$\,$3).
The atomic temperatures in this measurement range between $500\:$nK and $1.1\:\mu$K, well above the critical temperatures for Bose-Einstein
condensation. The atomic densities can therefore be described using a Maxwell-Boltzmann distribution.
Assuming a simple power-law dependence of the form $\bar{\Gamma}_\textrm{ion}\propto I_\textrm{L}^\alpha$ we obtain the best fit
using an exponent $\alpha=1.5(1)$ (solid green line in Fig.$\,$\ref{fig:4}a). This fit is between a linear and
a quadratic intensity dependence (dashed red and blue lines, respectively).
Thus, at least two of the three transitions composing the ionization process are partially saturated at the typical intensities used.

\noindent In order to better circumvent possible density
variations of the atomic cloud induced by changes in laser
intensity, we employ a further method which enables us to vary the
intensity with negligible effects on the atomic sample. We achieve
this by keeping the time-averaged intensity $\langle
I_\textrm{L}\rangle$ constant and comparing the ion production
rates within a continuous dipole trap and a ``chopped" dipole trap in
which the intensity is rapidly switched between $0$ and
$2I_\textrm{L}$. In both cases the trap is operated at an
intensity $\langle I_\textrm{L}\rangle\approx 15
\:\text{kW}/\text{cm}^2$. In the ``chopped" configuration the
intensity is switched at a frequency of $100\:\textrm{kHz}$ so
that the atoms are exposed to the light for $5\:\mu\textrm{s}$
followed by $5\:\mu\textrm{s}$ without light. It should be noted
that molecules formed in the ``dark" period with sufficiently high
kinetic energies may leave the central trapping region before the
laser light is switched back on. They are then lost for our REMPI
detection. Taking into account the molecular velocity and the
transverse extensions of the laser beams we can estimate that this
potential loss mechanism leads to errors of less than 30\%, even
at the highest binding energies relevant to this work
($E_\textrm{b}\approx h\times 750\:$GHz, see below). We did not
observe evidence of such losses experimentally. Investigations
were made by changing the chopping frequency. We define $R$ as
ratio of the ion production rates in the ``chopped" and the
continuous trap configuration. Fig.$\,$\ref{fig:4}b shows the
results of these measurements for various laser frequencies
$\nu_\text{L}$. We find a value $R\approx 1.5$ for off-resonant
frequency settings $\nu_\text{L}- \nu_\text{L}^0<
0.4\:\textrm{GHz}$, in good agreement with the result presented in
Fig.$\,$\ref{fig:4}a. When scanning the laser onto resonance at
$\nu_\text{L}- \nu_\text{L}^0 \approx 0.45\:$GHz (see also
Fig.$\,$\ref{fig:3}b) we obtain $R\approx 1$. This result
indicates a linear intensity dependence of the REMPI process in
the resonant case, which is explained by the saturation of two of
the three molecular transitions involved. It is known that
transitions into the ionization continuum (photon III, see
Fig.$\,$1e) will not saturate under the present experimental
conditions. This means that the excitation pathway via photon I
and II must be saturated and therefore both close to resonance.

However, given the wavelength range of about
1064.5$\pm 0.15\,$nm, an inspection of the level structure shows
that photon I can only resonantly drive three different
transitions which connect vibrational levels in states $X$ and $a$
to vibrational levels in states $A, b$ and $c$ (see Fig.$\,$1e).
Spectroscopic details for these transitions and the corresponding
vibrational levels  are given in the Methods section and in Fig.$\,$5.
From recent spectroscopic studies\cite{Strauss2010,
Takekoshi2011, Drozdova2012} and additional
measurements in our lab\cite{Rbexp} the level structure
of all relevant levels of the $X, a,
A, b,$ and $c$ states is well known. The absolute precision of most of the
level energies is far better than 1$\:$GHz for low rotational
quantum numbers $J$.

In the experimental data (Fig.$\,$\ref{fig:3}a) the central region
from $\nu_L - \nu^0_L = -6$ to 7$\:$GHz is marked by several
prominent resonances that are significantly stronger than those
observed throughout the rest of the spectrum. These resonance
peaks can be explained by transitions from the $X$ ground state to
$A$ and $b$ states. The prominence of these singlet transitions is
explained by the near degeneracy of levels due to small hyperfine
splittings. Indeed, by analyzing these strong resonances with
regard to line splittings and intensities it was possible to
consistently assign rotational ladders for total nuclear spin
quantum numbers $I=1,2,3$ for the transition $X (v = 115)
\rightarrow A (v' = 68)$. The starting point of the rotational
ladder for $I = 2$ was fixed by previous spectroscopic
measurements\cite{Rbexp}. At frequencies $\nu_L - \nu^0_L \gtrsim
2\:$GHz additional strong lines appear that we attribute to the
$X\:(v=109) \rightarrow b\:(v'=72)$ transition. The fact that we
observe $X$ state molecules with $I = 1, 2, 3$ is interesting
because for $I=1,3$ the total parity of the molecule is negative,
while for $I=0,2$ it is positive. However, a two-body collision
state of our spin polarized Rb atoms necessarily has positive
total parity due to symmetry arguments and a photoassociation
pathway would lead to ground state levels with positive parity.
The observed production of
molecules with negative total parity must then be a three-body
collision effect.

We now consider the role of secondary atom-molecule
collisions which would change the product distribution due to
molecular relaxation. Two aspects are of importance: 1)
depopulation of detected molecular levels and 2) population of
detected molecular levels via relaxation from more weakly bound
states.
In our experiments reported here we detect molecules that are
formed in states with binding energies on the order of hundreds of
GHz. These molecules leave the reaction with kinetic energies of
several K$\times k_\text{B}$. At these energies the rate
coefficients for depopulating atom-molecule collisions are small
(see e.g. ref.$\,$\cite{Simoni2006}) and the collision probability before the molecule
is either ionized or has
left the trap is below 1$\,$\%. \\
For the population processes, we can estimate an upper bound for
rate coefficients by assuming recombination to occur only into the
most weakly bound state with a binding energy of $24\,$MHz$\times
h$. In this case subsequent atom-molecule collision rates will be
roughly comparable to those expected in the ultracold limit. At
typical rate coefficients of $10^{-10}\,\text{cm}^3/\text{s}\;$
(see refs.
\cite{Mukaiyama2004,Staanum2006,Zahzam2006,Quemener2007}) and the
atomic densities $n_\text{0}\sim1\times 10^{13}\:\textrm{cm}^{-3}$
used in the measurement shown in Fig.$\,$\ref{fig:2}, the
collision probability before the molecule leaves the atom cloud is
around 5$\,$\%. This small probability grows linearly with density
so that the density dependence of the ion production rate should
show a significant cubic contribution if secondary collisions were involved (as
expected for this effective four-body process). This is
inconsistent with the data and thus indicates that the population
that we detect is not significantly altered by secondary collisions.

\begin{figure}
\begin{center}
\vspace{6pt}
\includegraphics[width = 0.45\textwidth]{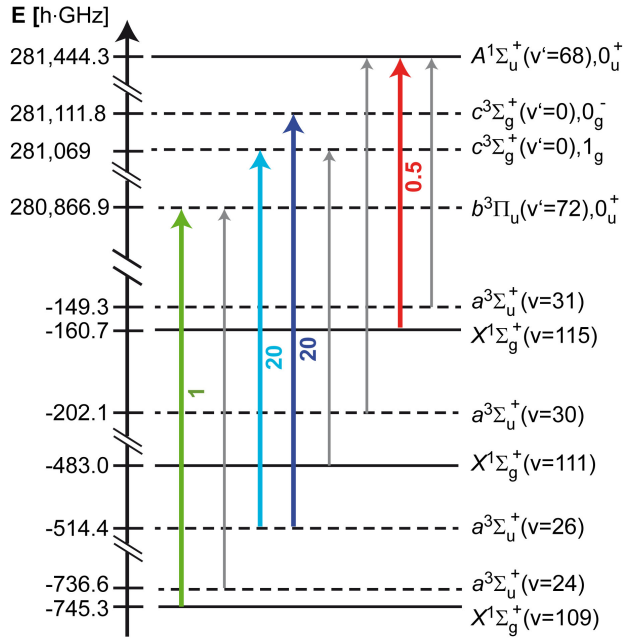}
\end{center}
\caption{\bf{Overview over relevant molecular levels and transitions.}
The vertical axis denotes the energy $E_b$ of the energetically lowest levels
of each vibrational manifold with respect to the 5s5s asymptote. Colored thick arrows represent molecular transitions relevant to the spectrum of
Fig.$\,$\ref{fig:3}. The expected relative strengths of these
transitions are also given. Grey arrows mark transitions that occur in
the relevant spectral region but are so weak that they can be neglected
(for further spectroscopic details see Methods).
We identify three main molecular transitions
for the initial step of the REMPI process. (1) Molecules in
the $v=26$ vibrational level of the $a^3\Sigma_{u}^+$ potential
are excited to the $v'=0$ level of the $c^3\Sigma_{g}^+$ potential.
This level is split into a $1_{g}$ and a $0_{g}^-$ component. (2)
Excitation from $X^1\Sigma_{g}^+ (v=115)$ to $A^1\Sigma_u^+
(v'=68)$. (3) Excitation from $X^1\Sigma_{g}^+ (v=109)$ to
$b^3\Pi_u (v'=72)$. Transition (3) becomes possible through the
strong spin-orbit coupling of the $A$ and $b$ states.}
\label{fig:5}
\end{figure}

We can roughly estimate the range of
molecular rotation $J$ of the populated levels in the ground state.
The strong isolated lines that we have assigned to the $X \: (v = 115) \rightarrow
A \: (v' = 68) $ transition are all contained within a relatively
small spectral region ($|\nu_L - \nu^0_L| < 6\:$GHz)
and are explained by rotational quantum numbers $J\leq 7$. Population
of higher
rotational quantum numbers would result in a continuation of the strong
resonance lines stretching to transition frequencies beyond
$\nu_L - \nu^0_L=10\:$GHz, which we do not observe. Similarly,
if only rotational quantum numbers $J\leq 5$ were populated, a spectrum
 would result which does not have enough lines to explain the data. Thus, we can
roughly set the limits on the molecular rotation to $J\leq 7$, a value that is
also consistent with our observations of the spread
of the transitions $X \rightarrow b$ and $a \rightarrow c$ (see Fig.$\,$\ref{fig:5}).
Finding quantum numbers as high as $J = 7$ is remarkable because the three-body collisions
at $\mu$K temperatures clearly take place in an $s$-wave regime, i.e. at vanishing rotational
angular momentum. Hence, one could expect to produce $X$ state molecules dominantly at $J = 0$,
which, however, we do not observe.

Despite the limited spectral range covered by our measurements, we can
already estimate the number of molecular vibrational levels populated
in the recombination events. From the three states $X \: (v=109)$, $a\: (v=26)$ and $X\: (v = 115)$ that
we can observe within our wavelength range, all deliver comparable signals in the spectrum of
Fig.$\,$\ref{fig:3}. This suggests that at least all vibrational states more weakly bound than $X \: (v=109)$
should be populated, a total
of 38 vibrational levels (counting both singlet and triplet states).
This is a significant fraction of the 169 existing levels of the $X$ and $a$ states,
although restricted to a comparatively small range of binding energies.

In conclusion, we have carried out a first, detailed experimental
study of the molecular reaction products after three-body
recombination of ultracold Rb atoms. We use a high-power,
narrow-linewidth laser to state-selectively ionize the
produced molecules in a REMPI process. Subsequently, these ions are
trapped in an ion trap and detected with very high sensitivity and
negligible background. An analysis of the ionization spectrum
allows us to identify population of several vibrational quantum
levels indicating that the recombination events result in a
fairly broad and uniform population distribution. We
conjecture that dozens of vibrational levels are populated in
total. Molecules are produced both in $X^1\Sigma_g^+$ as well as
$a^3\Sigma_u^+$, with negative and positive total parity, various total
nuclear spins and rotational quantum numbers reaching $J \leq 7$.
Our work represents a first experimental step towards a detailed understanding on
how the reaction channels in three-body recombination  are populated.
A full understanding will clearly require further experimental and theoretical efforts. On the experimental side the scanning range has to be increased and it could be advantageous to switch to a two-color REMPI scheme in the future.
Such studies may finally pave the way to a comprehensive understanding
of three-body recombination, which includes the details
of the final products.\\
\noindent Reaching beyond the scope of three-body recombination,
the great sensitivity of our detection scheme has enabled us to state-selectively
probe single molecules that are produced at rates of only a few Hz.
We thereby demonstrate a novel scheme for precision molecular spectroscopy
in extremely dilute ensembles.

\parskip 12pt
\begin{methods}

\subsection{Dipole trap and REMPI configuration.}
The crossed dipole trap is composed of a horizontal and a vertical
beam focussed to beam waists of $\mathtt{\sim}90\:\mu\textrm{m}$
and $\mathtt{\sim}150\:\mu\textrm{m}$, respectively. It is
positioned onto the nodal line of the radiofrequency field of the
linear Paul trap with $\mu \text{m}$ precision. The two trap
centers are separated by about $300\:\mu\textrm{m}$ along the axis
of the Paul trap (see Fig. 1d). In a typical configuration, the
trap frequencies of the dipole trap are
$(175,230,80)\,\textrm{Hz}$ resulting in atom cloud radii of
about $(6,7,16)\,\mu\textrm{m}$. The short-term frequency
stability of the dipole trap laser source is on the order of
$1\:\text{kHz}$ and it is stabilized against thermal drifts to
achieve long-term stability of a few MHz. The two beams of the
dipole trap are mutually detuned by $160\:\textrm{MHz}$ to avoid
interference effects in the optical trap. Consequently, two
frequencies are in principle available to drive the REMPI process.
However, the intensity of the horizontal beam is 4 times
larger than the one of the vertical beam and we have not directly
observed a corresponding doubling of lines.
Further details on the
atom-ion apparatus are given in ref$\,$\cite{Smi2012}.

\subsection{Paul trap configuration.} The linear Paul trap is driven at a radiofrequency of $4.17\:\textrm{MHz}$
and an amplitude of about $500\:\textrm{V}$ resulting in radial
confinement with trap frequencies of
$(\omega_\textrm{x,Ba},\omega_\textrm{y,Ba})=2\pi\times(220,230)\,\textrm{kHz}$
for a $^{138}\textrm{Ba}^+$ ion. Axial confinement is achieved by
applying static voltages to two endcap electrodes yielding
$\omega_\textrm{z,Ba}=2\pi\times40.2\:\textrm{kHz}$. The trap
frequencies for "dark" Rb$_2^+$ and Rb$^+$  ions produced in the
REMPI processes are
$(m_\textrm{Ba}/m_\textrm{dark}\times\omega_\textrm{x,Ba},
m_\textrm{Ba}/m_\textrm{dark}\times\omega_\textrm{y,Ba},\sqrt{m_\textrm{Ba}/m_\textrm{dark}}\times\omega_\textrm{z,Ba})$
where $m_\textrm{Ba}$ and $m_\textrm{dark}$ denote the mass of the Ba$^+$ ion and the dark ion, respectively. The depth of the
Paul trap depends on the ionic mass and exceeds $2\:\textrm{eV}$ for all ionic species relevant to this work.

\subsection{Ion detection methods.} We employ two methods to detect
Rb$_2^+$ and Rb$^+$ ions both of which are not amenable to
fluorescence detection. In the first of these methods we use a
single trapped and laser-cooled $^{138}$Ba$^+$ ion as a probe. By recording
its position and trapping frequencies in small ion strings with up
to 4 ions we detect both the number and the masses of the ions
following each REMPI process (see also\cite{Smi2010}). The second
method is based on measuring the number of ions in the Paul trap
by immersing them into an atom cloud and recording the ion-induced
atom loss after a hold time of $2\:\textrm{s}$ (see
also\cite{Harter2012}). During this detection scheme, we take care to
suppress further generation of ions by working with small and
dilute atomic clouds and by detuning the REMPI laser from
resonance. Both methods are background-free
in the sense that no ions are captured on timescales of days
in the absence of the atom cloud. Further information on both detection
methods is given in the Supplementary Information.

\subsection{Spectroscopic details.}
Spin-orbit and effective spin-spin coupling in the $A$,
$b$, and $c$ states lead to Hund's case c coupling where the relevant
levels of states $A$ and $b$ have $0_u$ symmetry while the levels
of state $c^3\Sigma_g^+$ are grouped into $0_g^-$ and $1_g$
components. The level structure of the $0_{u}^+$ states is quite simple as
it is dominated by rotational splittings. Typical rotational constants for the electronically
excited states are on the order of 400$\:$MHz, for the weakly bound $X$ and $a$ states
they are around 100-150$\:$MHz.

Figure 5 shows the relevant optical transitions between the $X, a$
states and the $A, b, c$ states  in our experiment. For the given
expected relative strengths of these transitions, we only consider
Franck-Condon factors and the mixing of singlet and triplet
states, while electronic transition moments are ignored. The
colored arrows correspond to transitions with large enough
Franck-Condon factors (typ. $10^{-2} ... 10^{-3})$  so that at
laser powers of $\approx 10^4\,$W/cm$^2$ resonant transitions can
be well saturated. Transitions marked with grey arrows can be
neglected due to weak transition strengths, resulting from small
Franck-Condon factors or dipole selection rules.

\end{methods}

The authors would like to thank Stefan Schmid and Andreas Brunner
for support during early stages of the experiment and Olivier
Dulieu, Brett Esry, Jose d'Incao, William Stwalley, Ulrich Heinzmann, Jeremy
Hutson, Pavel Soldan, Thomas Bergeman and Anastasia Drozdova for valuable
information and fruitful discussions. This work was supported by
the German Research Foundation DFG within the SFB/TRR21.

\bibliography{denseionization_submission}

\cleardoublepage

\begin{center}
\textbf{\Huge{Supplementary Information}}
\end{center}
\setcounter{figure}{0}

In this Supplementary Information we describe two methods that we employ to detect small numbers 
of Rb$_2^+$ and Rb$^+$ ions in our linear Paul trap.
\\

\textbf{\Large{Method 1}}

\begin{figure}
\begin{center}
\vspace{6pt}
\includegraphics[width = 0.40\textwidth]{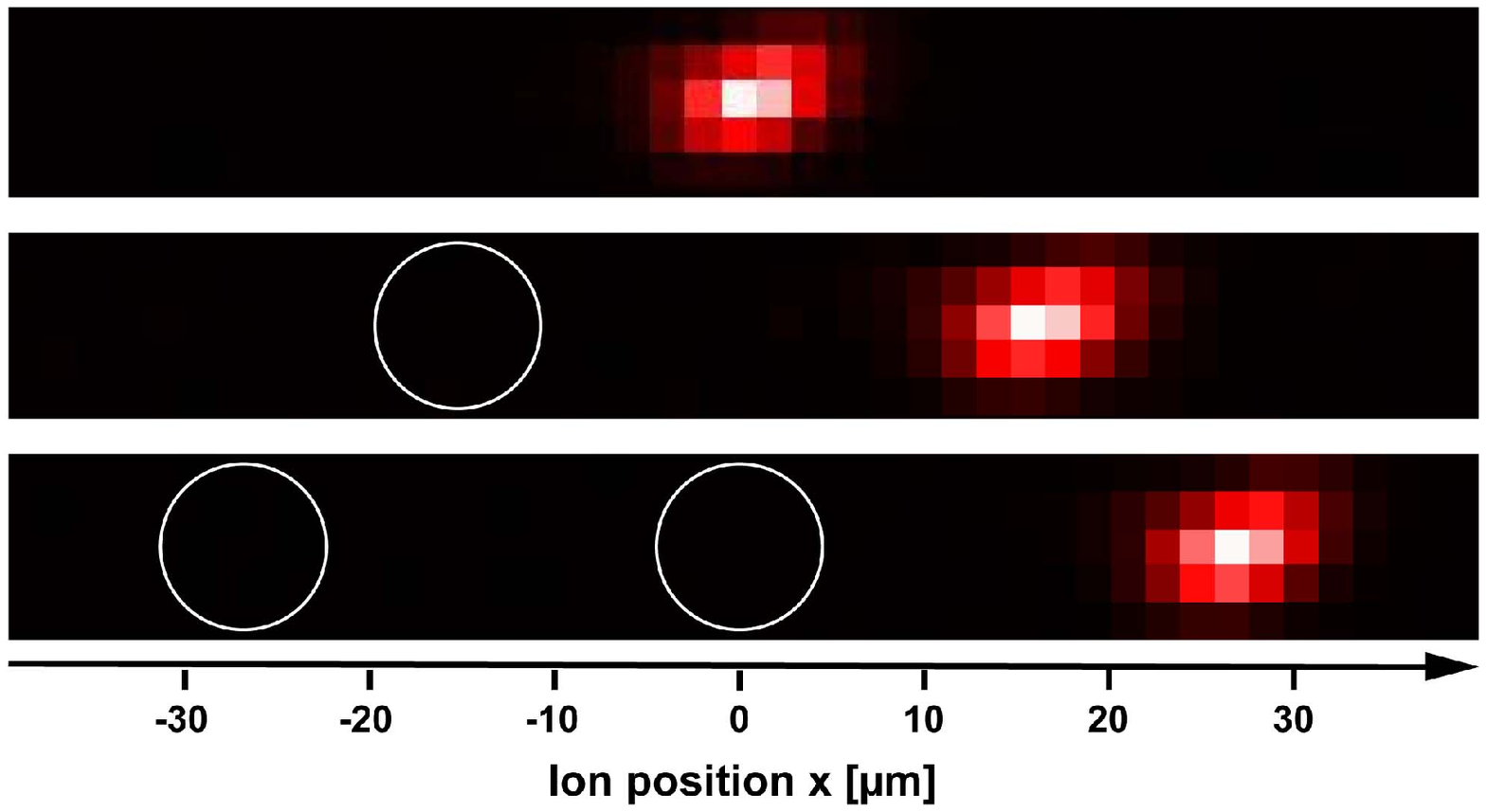}
\end{center}
\caption{\capbf{Ion detection using a $^{138}$Ba$^+$-ion.} Positional shifts of the fluorescence of the Ba$^+$-ion 
and measurements of the trap oscillation frequencies allow us to perform mass-sensitive detection of up to 
three "dark" ions in the trap.} 
\label{fig:sup1}
\end{figure}

\noindent To implement our first ion detection method allowing mass-sensitive detection of "dark" ions we rely 
on the presence of a single "bright" ion in the trap. Information on additional ions
can be extracted from its fluorescence position.
When using this method, our experimental procedure begins with the loading of a single $^{138}$Ba$^+$-ion into our linear Paul trap. 
We laser-cool the ion and image its fluorescence light onto an electron-multiplying 
charge-coupled device camera. This enables us to determine the position of the trap center to better 
than $100\:\textrm{nm}$. The ion is confined at radial and axial trapping frequencies 
$\omega_\textrm{\,r,Ba} \approx 2\pi\times 220\:\textrm{kHz}$ and $\omega_\textrm{ax,Ba} \approx 2\pi\times 
40.2\:\textrm{kHz}$ and typically remains trapped on timescales of days. Next, we prepare an ultracold atomic sample 
in the crossed dipole trap. At typical atomic temperatures of about $700\:$nK
the atom cloud has radial and axial extensions of about $7\:\mu$m and $15\:\mu$m and is thus much smaller than
the trapping volume of our Paul trap. To avoid atom-ion collisions we 
shift the Ba$^+$-ion by about $300\:\mu$m with respect to the atom cloud before the atomic sample arrives in the 
Paul trap. The shifting is performed along the axis of the trap by lowering the voltage on one of the endcap electrodes. 
Additionally, we completely extinguish all resonant laser light so that the atoms are only subjected to the light of the dipole
trap. The atomic sample is moved into the center of the radial trapping potential of the Paul trap  and is typically held 
at this position for a time $\tau_{\,\textrm{hold}}\approx 10\,\textrm{s}$. Despite the 
axial offset from the center of the Paul trap, the atom cloud at this position is fully localized within the trapping
volume of the Paul trap. After the hold time the sample is detected using absorption imaging. Subsequently, the ion 
cooling beams are switched back on for fluorescence detection of the Ba$^+$-ion.\\
The presence of a second ion in the trap leads to positional shifts of the $^{138}$Ba$^+$-ion by distances on the order 
of $10\:\mu$m (see Fig.$\,$\ref{fig:sup1}). We make use of the mass-dependent trap frequencies of the Paul trap to gain 
information on the ion species 
trapped. In a two-ion Coulomb crystal composed of a Ba$^+$-ion and a dark ion, the axial center-of-mass frequency 
$\omega_\textrm{\,ax,2ion}$ shifts with respect to $\omega_\textrm{ax,Ba}$ depending on the mass of the dark ion 
$m_\textrm{dark}$ \cite{Morigi2001}. We measure $\omega_\textrm{\,ax,2ion}$ by modulating the trap drive at 
frequencies $\omega_\textrm{mod}$ and by monitoring the induced axial oscillation of the Ba$^+$-ion, visible as a blurring 
of the fluorescence signal. In this way, after each ion trapping event, we identify a resonance either 
at $\omega_\textrm{mod} \approx 2 \pi \times 44 \:\textrm{kHz}$ or $\omega_\textrm{mod} \approx 2 \pi \times 38 \:\textrm{kHz}$ 
corresponding to $m_\textrm{dark}=87\:\textrm{u}$ and $m_\textrm{dark}=174\:\textrm{u}$, respectively (see table~\ref{table1}).\\

We have expanded this method for ion strings with up to four ions including the Ba$^+$-ion. For this purpose,
we perform the following step-by-step analysis.
\begin{enumerate}
\item The position $x$ of the Ba$^+$-ion with respect to the trap center is detected. If $x\neq 0$, the value
of $x$ allows us to directly determine the total number of ions in the string. 
\item If $x=0$ we need to distinguish between a single Ba$^+$ ion and a three-ion string with Ba$^+$ at its center.
This is done by modulating the trap drive at $\omega_\textrm{ax,Ba}$, thereby only exciting the Ba$^+$ ion 
if no further ions are present.
\item We destructively detect the Rb$^+$ ions by modulating the trap drive on a $5\:\text{kHz}$ wide band around
$2\times \omega_\textrm{r,Rb}/(2\pi)=691\:\textrm{kHz}$. This selectively removes only Rb$^+$ ions from the string making
use of the relatively weak inter-ionic coupling when exciting the ions radially. 
\item Steps 1. and 2. are repeated to detect the number of remaining ions. 
\item The Rb$_2^+$ ions are destructively detected via modulation around $2\times \omega_\textrm{r,Rb2}/(2\pi)=341\:\textrm{kHz}$.
\end{enumerate}

\begin{table}
\centering
\caption{\capbf{Trap oscillation frequencies of two-ion crystals}}

\footnotesize\rm
\begin{tabular}{c | r | r}
Ion species & $\omega_\textrm{\,ax,2ion}/2\pi$ [kHz] & $\omega_\textrm{\,r}/2\pi$ [kHz]\\
\hline
$^{138}$Ba$^+$ and $^{138}$Ba$^+$  & 40.2   & 220.0\\
$^{138}$Ba$^+$ and $^{87}$Rb$^+$   & 44.0   & 345.3\\
$^{138}$Ba$^+$ and $^{87}$Rb$_2^+$ & 37.7   & 170.7
\label{table1}
\end{tabular}
\end{table}

\ 
\\

\textbf{\Large{Method 2}}

\begin{figure}
\begin{center}
\vspace{6pt}
\includegraphics[width = 0.45\textwidth]{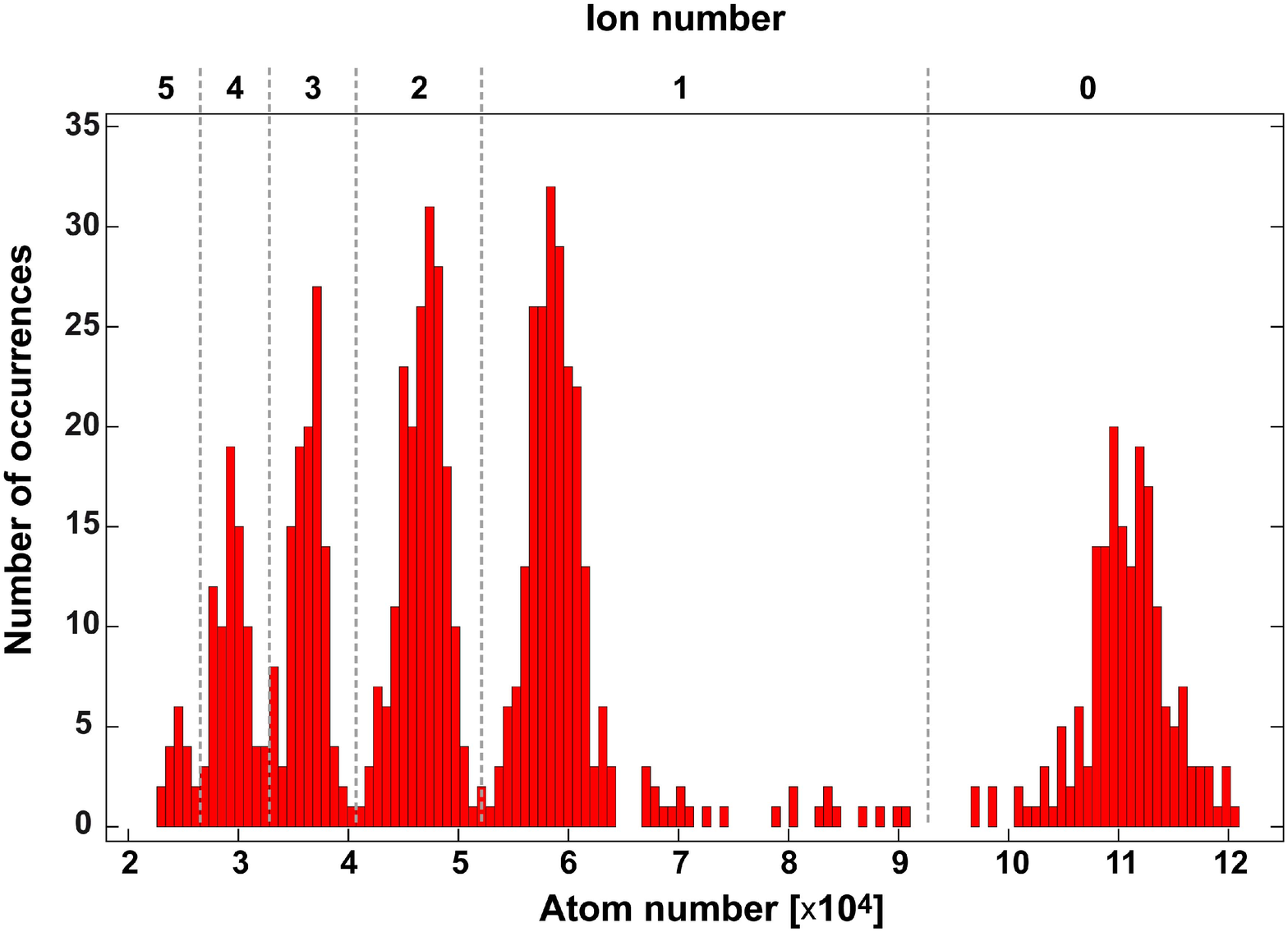}
\end{center}
\caption{\capbf{Ion detection method based on ion-induced atom loss.} We overlap an ultracold atom cloud
containing approximately 110,000 atoms with the center of the Paul trap. After an interaction time $\tau=2\,\textrm{s}$
we detect the ion-induced atom loss via absorption imaging of the atom cloud. The discrete number of trapped ions
is clearly reflected in the displayed histogram of atom numbers.} 
\label{fig:sup2}
\end{figure}

\noindent We have also developed a second ion detection method that does not require an ion fluorescence signal. Instead, 
the trapped ions are detected via their interaction with an atomic sample. 
For this purpose, we produce a comparatively small atom cloud containing about $1\times 10^5$ atoms at a density
of a few $10^{12}\:\textrm{cm}^{-3}$. In addition, we set the frequency of the dipole trap laser to an off-resonant 
value so that the production of additional ions during the ion probing procedure becomes extremely unlikely. We now fully
overlap the ion and atom traps for an interaction time $\tau_\textrm{\footnotesize{int}}=2\,\textrm{s}$. By applying an external electric field
of several V/m we set the ion excess micromotion energy to values on the order of tens of $k_\text{B}\times$mK \cite{Berkeland1998, Harter2012}.
 Consequently, if ions are present in the trap, strong atom losses occur due to elastic atom-ion collisions. Fig.\ref{fig:sup2} shows a
histogram of the atom numbers of the probe atom samples consisting of the outcome of about 1,000 experimental runs. 
The histogram displays several peaks which can be assigned to the discrete 
number of ions in the trap. Up to five ions were trapped simultaneously and detected with high fidelity. 
The atom loss rate increases nonlinearly with ion number mainly because the interionic repulsion prevents the ions from all occupying the 
trap center where the atomic density is maximal. While ion detection method 2 does not distinguish ionic masses, 
it has advantages in terms of experimental stability and does not require the trapping of ions 
amenable to laser cooling or other fluorescence based detection techniques. 

\end{document}